\documentclass[10pt,conference]{IEEEtran}
\usepackage{epsfig,setspace,amsmath,epsf,amssymb,bm,theorem,cite,graphicx, epstopdf,algorithm,float,color,mathtools, authblk,physics}
\usepackage[table,xcdraw]{xcolor}
\usepackage{subcaption}
\usepackage{algorithm,algorithmic}
\usepackage{dsfont}
\usepackage[short,c2]{optidef}

\DeclareMathOperator*{\argmax}{arg\,max}

\newtheorem{theorem}{Theorem}
\newtheorem{corollary}{Corollary}

\newtheorem{remark}{Remark}
\newtheorem{lemma}{Lemma}

\newcommand{\e}{{\mathbb{E}}}

\IEEEoverridecommandlockouts
\allowdisplaybreaks

\begin{document}

\title{Multi-Stage Structured Estimators for \\ Information Freshness}

\author[1]{Sahan Liyanaarachchi}
\author[1]{Sennur Ulukus}
\author[2]{Nail Akar}

\affil[1]{\normalsize University of Maryland, College Park, MD, USA}
\affil[2]{\normalsize Bilkent University, Ankara, T\"{u}rkiye}

\maketitle
\begin{abstract}
Most of the contemporary literature on information freshness solely focuses on the analysis of freshness for martingale estimators, which simply use the most recently received update as the current estimate. While martingale estimators are easier to analyze, they are far from optimal, especially in pull-based update systems, where maximum aposteriori probability (MAP) estimators are known to be optimal, but are analytically challenging. In this work, we introduce a new class of estimators called $p$-MAP estimators, which enable us to model the MAP estimator as a piecewise constant function with finitely many stages, bringing us closer to a full characterization of the MAP estimators when modeling information freshness.
\end{abstract}

\section{Introduction}
With the dawn of 6G communications, spanning from the modern day vehicular networks to the outskirts of cislunar communications, timeliness of information has become a crucial feature that must be integrated into any contemporary communication infrastructure \cite{AoI_self_drive, yuan_towards,sahan-cislunar}. Henceforth, a variety of metrics, such as, age of information (AoI) \cite{yates2020age,age1,age2}, age of incorrect information (AoII) \cite{AoII2019,AoII_Markov} and binary freshness (BF) \cite{melih_BF_cache, melih_IF_CUS, melih_BF_Inf, melih_BF_gossip} have been developed to evaluate and quantify the timeliness of information in such freshness-critical systems. Among them, BF is the most widely used metric for modeling the information freshness when monitoring Markov sources, due to its direct relation to error probability. Let BF be denoted by $\Delta(t)$. Then, $\Delta(t)=1$, if the estimator is in sync with the source and zero, otherwise. Now, the mean binary freshness (MBF), denoted by $\e[\Delta]$, can be expressed as,
\begin{align}
    \e[\Delta]=\limsup_{T\to\infty}\frac{1}{T}\e\left[\int_0^T\mathds{1}\{X(t)=\hat{X}(t)\}\right],
\end{align}
where $\mathds{1}\{\cdot\}$ is the indicator function. 

Most of the works considering BF has been restricted to the analysis of freshness under martingale estimators, which estimate the state of the source as being equal to the latest received sample, mainly due to their simplicity. Despite their simplicity, martingale estimators can have an adverse effect on the freshness of information, especially in pull-based update systems with a limited sampling budget. For such systems, if the martingale estimator lands on a less probable state, then the system is forced to retain this state as its estimate until the next sampling instance. For such systems, MAP estimates are ideal, but can be analytically cumbersome as they can be unstable  (infinitely oscillating) in some cases.

\begin{figure}
    \centering
    \includegraphics[width=\linewidth]{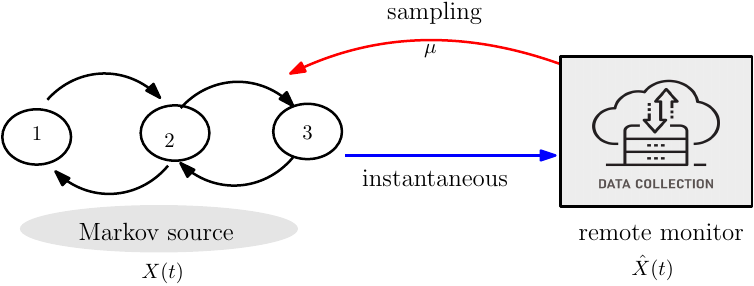}
    \caption{Query-based sampling of a Markov source.}
    \label{fig:sys_model}
\end{figure}

Motivated by this, the previous work in \cite{erlang_est} introduces a new structured estimator known as the $\tau$-MAP estimator, which can seamlessly shift between the martingale and MAP estimators. \cite{erlang_est} considers the problem of monitoring a continuous time Markov chain (CTMC) using  a query-based sampling procedure (see Fig.~\ref{fig:sys_model}), which samples the CTMC at time intervals following an exponential distribution. \cite{erlang_est} shows that the $\tau$-MAP estimators have a significant freshness gain over martingale estimators. However, the $\tau$-MAP estimators do not fully characterize the MAP estimator and the class of CTMCs for which it is applicable is restrictive. 

In this work, we improve upon the work in \cite{erlang_est} by introducing a new class of estimators coined as $p$-MAP estimators, which models the MAP estimator as a piecewise constant function with a fixed finite set of transition points (stages). We show that the $p$-MAP estimator is consistent with the MAP estimator for time-reversible CTMCs and provide analytical expressions for its freshness. Additionally, we show that by employing a state-dependent sampling policy, we can further achieve significant gains in terms of freshness. The proofs will be available in the journal version of this paper. 

\section{Related Work}
The problem of query-based sampling of CTMCs was introduced in \cite{nail_QS}, which considered maximizing several variants of BF using the martingale estimator. The work in \cite{Markov_machines} looks into the problem of monitoring multiple Markov machines (MM) using a query-based sampling procedure, where they consider an estimate-dependent decision process that influences the state of the MM. In \cite{Markov_machines} too, the martingale estimator was used to govern the decision process. This work was later extended in \cite{revMax}, where the authors used the MAP estimator of the system to improve the decision process so as to maximize the revenue. On a separate avenue, the work in \cite{ismail_map} looks into the problem of minimizing AoII in a pull-based update system by considering the MAP estimate. However, they resort to reinforcement learning (RL) techniques to overcome the analytical challenges posed by the MAP estimator when modeling a decision process.

The closest to our work is \cite{erlang_est}, which introduced  the $\tau$-MAP estimator as a surrogate for the MAP estimator when monitoring CTMCs whose stationary distribution has a unique maximum. For such CMTCs, once the age of the estimate exceeds a certain threshold, the MAP estimator will remain constant regardless of the initial state. Hence, \cite{erlang_est} poses the $\tau$-MAP estimator as a two-stage approximation of the MAP estimator, where initially, it will be equal to the martingale estimator and once the age of the estimator exceeds the threshold, it will shift to the MAP estimate of the system. The analysis in \cite{erlang_est} relies on constructing an Erlang chain for modeling  deterministic thresholds in CTMCs, and hence, is not easily extendable if a more refined approximation for the MAP estimate is to be used. In our work, we model the MAP estimator as a piecewise constant function, a multi-stage approximation, to the MAP estimator.

\section{System Model}
Let $X(t)\in\mathcal{S}=\{1,2,\dots,S\}$ be a finite, irreducible CTMC with $S$ states and let $Q=\{q_{ij}\}_{i,j\in \mathcal{S}}$ be its generator matrix, where $q_{ij}$ denotes the transition rate from state $i$ to $j$ with $q_{ii}=-\sum_{j\neq i}q_{ij}$. Let $\hat{X}(t)$ be the estimate of this CTMC maintained at the remote monitor. This CTMC is monitored via a query-based sampling procedure, where the remote monitor sends queries to the CTMC at time intervals following a sample-dependent exponential distribution. In particular, if the latest sample indicated that the CTMC was in state $i$, then it will be sampled at a rate of $\mu_i$. We assume that once sampled, the new sample is transmitted instantaneously to the remote monitor. Hence, $X(t)=\hat{X}(t)$ at sampling instances.

Let $\bm{\pi}=\{\pi_1,\pi_2,\dots,\pi_S\}$ be the stationary distribution of the CTMC, where $\bm\pi Q=\bm{0},$ and let $\Pi$ be a diagonal matrix whose diagonal is $\bm{\pi}$. We say that $X(t)$ is \emph{time-reversible} if it satisfies the detailed balanced equations $\pi_i q_{ij}= \pi_jq_{ji}$, $\forall i,j\in \mathcal{S}$  and $i\neq j$. For time-reversible CTMCs, we have that $\Pi^{\frac{1}{2}}Q\Pi^{-\frac{1}{2}}$ is symmetric and hence can be diagonalized as $U D U^T$, where $D$ is a diagonal matrix whose diagonal entries are $\{d_1,-d_2,\dots,-d_S\}$, with $d_1=0$ and $d_i>0$ for $i\geq 2$\cite{gallager}. Let $P(t)$ be the transition matrix of $X(t)$, where $P_{ij}(t)$ is the probability that $X(t)=j$ given $X(0)=0$. 

Now, by using Lemma \ref{LEM:PIJ}, we can fully characterize $P_{ij}(t)$.

\begin{lemma}\label{LEM:PIJ}
    For time-reversible CTMCs, the elements of the transition matrix are given as follows,
    \begin{align}
        P_{ij}(t)=\sqrt{\frac{\pi_k}{\pi_i}}\sum_{k=1}^Su_{ki}u_{kj}e^{-d_kt},
    \end{align}
    where $u_{ij}$ are the elements of matrix $U^T$ that diagonalizes $Q$.
\end{lemma}

The proof of Lemma \ref{LEM:PIJ} directly follows from $P(t)=e^{Qt}$. 

\begin{figure}
\captionsetup[subfigure]{aboveskip=1pt,belowskip=1pt}
    \centering
    \begin{subfigure}[b]{0.4\textwidth}
        \centering
        \includegraphics[width=0.8\textwidth]{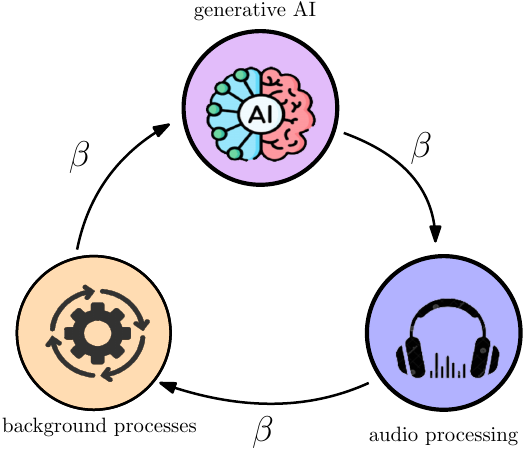}
        \medskip
        \caption{Non time-reversible CTMC modeling a CPU.}
        \bigskip
        \label{fig:non_tr_ctmc}
    \end{subfigure}
    \begin{subfigure}[b]{0.4\textwidth}
        \centering
        \includegraphics[width=\textwidth]{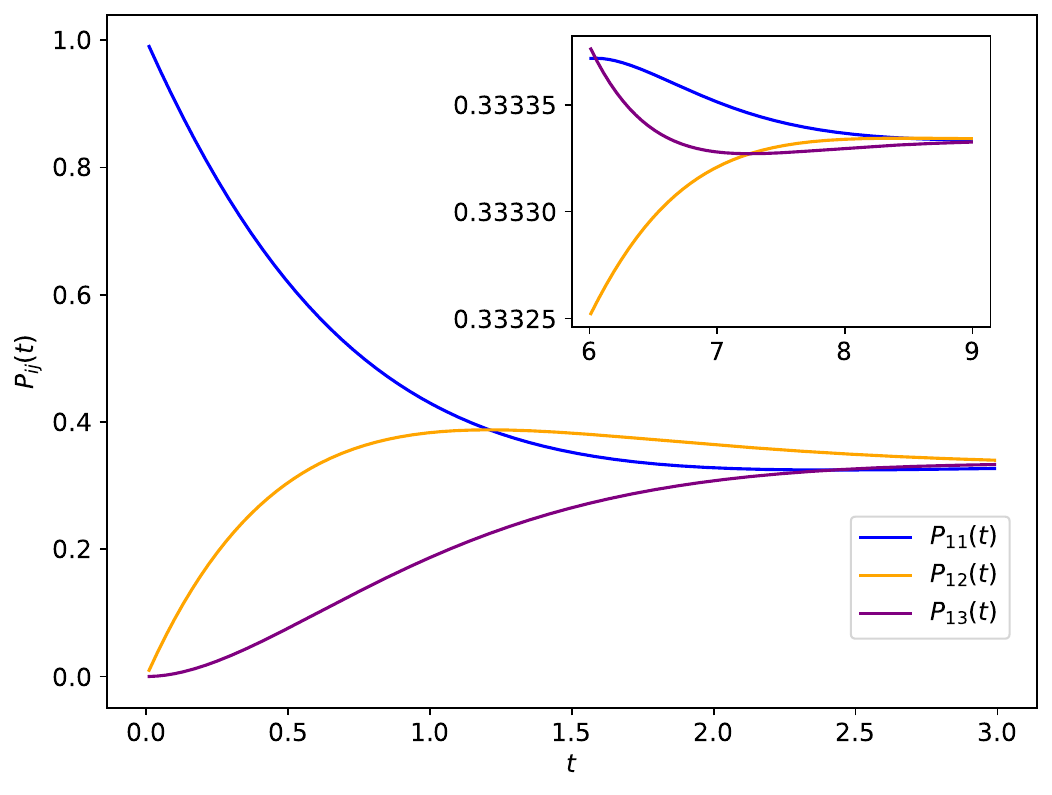}
        \caption{Transition probabilities.}
        \label{fig:osc}
    \end{subfigure}
    \caption{Infinitely oscillating transition probabilities for a non time-reversible CTMC which has no unique stationary maximum. Here, $\bm\pi=\left\{\frac{1}{3},\frac{1}{3},\frac{1}{3}\right\}$. Since all $\pi_i$, $i=1, 2, 3$ are equal, there is no unique $\argmax_i{\pi_i}$ here.}
    \label{fig:map_osc}
\end{figure}

Lemma \ref{LEM:PIJ} uncovers an important property for time-reversible CTMCs. That is, $P_{ij}(t)$ is a linear combination of exponential functions of the form $e^{-d_kt}$, and so are their differences, i.e., $f^i_{jk}(t)=P_{ij}(t)-P_{ik}(t)$, for $j\neq k$. Hence, by the virtue of repeated application of  Rolle's theorem, it follows that the function $f^i_{jk}(t)$ has only finitely many roots. Thus, there is only finitely many intervals in which either $P_{ij}(t)$ or $P_{ik}(t)$ is strictly better than the other. Therefore, since the state space $\mathcal{S}$ is finite, we can infer that the MAP estimate of $X(t)$, given an initial state $i$, is a piecewise constant function with finitely many transition points. This property holds for any CTMC whose generator matrix is entirely composed of real eigenvalues. Time reversibility ensures that all the eigenvalues are real. However, this statement is not always true for non time-reversible CTMCs, an example of which is shown in Fig.~\ref{fig:map_osc}. In Fig.~\ref{fig:map_osc}(b), the MAP estimator is infinitely oscillating where the oscillations are driven by the complex eigenvalues of the generator matrix. However, even if the generator matrix of the CTMC contains complex eigenvalues, \cite{erlang_est} showed that if the $\argmax_i{\pi_i}$ is unique, then there exists a $\tau^*>0$ such that the MAP estimate is equal to $\argmax_i{\pi_i}$ for all $t>\tau^*$, regardless of the initial state. We term such CTMCs to have a \emph{unique stationary maximum}. Motivated by these observations, we define the following set of structured estimators.

\subsection{Martingale Estimator}
Let $\hat{X}_M(t)$ denote the martingale estimator which simply retains the most recently received state of $X(t)$ as its estimate. Let us denote by $G(t)$ the time at which the most recent update was received. Then, $\hat{X}_M(t)$ can be defined as follows,
\begin{align}
    \hat{X}_M(t)=X(G(t)).
\end{align}

\subsection{$\tau$-MAP Estimator}
This is a simple modification of the $\tau$-MAP estimator which was introduced in \cite{erlang_est}, where we assume that if $\bm\pi$ does not have a unique maximum, then we choose the smallest index in the maximal set of $\bm\pi$. Let $\hat{X}_\tau(t)$ denote this estimator. Then, $\hat{X}_\tau(t)$ is defined as follows,
\begin{align}
    \hat{X}_\tau(t)=\begin{cases}
        \hat{X}_M(t), &\text{if}~\delta(t)\leq \tau,\\
        i^*, &\text{if}~\delta(t)>\tau,
    \end{cases}
\end{align}
where $i^*=\inf\{\argmax_i\pi_i\}$ and $\delta(t)=t-G(t)$ is the age of our estimate. That is, this is a two-stage estimator.

\subsection{$p$-MAP Estimator}
This is an extension of the $\tau$-MAP estimator, where we consider multiple intermediate transition stages for the estimator. Suppose our most recent sample indicates that the CTMC was in state $i$. Then, we will allow our estimator to evolve through at most $K_i$ stages before the next sampling instance, based on the age of the estimator. In here, a stage is an interval of time where our estimator remains constant. Let us  represent these intervals using the sequence of non-negative real numbers  $\{\tau_{i,0},\tau_{i,1},\dots,\tau_{i,K_i}\}$   with $\tau_{i,0}=0$, $\tau_{i,K_i}=\infty$ and $\tau_{i,k}\leq \tau_{i,k+1}$, where the interval $[\tau_{i,k},\tau_{i,k+1})$ represents the $k$th stage of our estimator if the most recently sampled state was $i$. As we wait for the next sample, the age of our estimator $\delta(t)$ increases. If $\delta(t)$ belongs to the interval $[\tau_{i,k},\tau_{i,k+1})$, i.e., the $k$th stage, then the estimator will assume the value $\Gamma_{i,k}$. Therefore, as the age of our estimator increases, the estimator will assume different values and will evolve through multiple stages. The number of stages, the duration of the stages, and the value the estimator assumed in each stage, all depend on the most recent sample, i.e., the most recent observation. 

Now, the $p$-MAP estimator, denoted by $\hat{X}_p(t)$, can be formally defined as follows,
\begin{align}
    \hat{X}_p(t)=\sum_{k=1}^{K_{i(t)}}\Gamma_{i(t),k}\mathds{1}\left\{\delta(t)\in \left[\tau_{i(t),k-1},\tau_{i(t),k}\right]\right\},
\end{align}
where  $\Gamma_{i,k}=\argmax_j\int_{\tau_{i,k-1}}^{\tau_{i,k}}P_{ij}(t)e^{-\mu_i t}\,\dd{t}$ for $k>1$ with $\Gamma_{i,1}=i$  and $i(t)=X(G(t))$. Sometimes we will use $i_k$ instead of $\Gamma_{i,k}$ for brevity. In here, $\Gamma_{i,k}$ is explicitly chosen to maximize the MBF metric which will be evident in Section \ref{sec:main_results}. We resort to this definition for the $p$-MAP estimator to facilitate the approximation of the MAP estimate with the desired number of intermediate stages. Moreover, this definition enables us to go beyond time reversibility and generalize to any CTMC (even ones with infinitely oscillating MAP estimators). However, in this work, we are mainly focused on time-reversible CTMCs since we can obtain closed-form expressions for them. In fact, for a time-reversible CTMC, if we replace $\tau_{i,k}$ with $\tau_{i,k}^*$ which is the $k$th transition point of the MAP estimate when $X(0)=i$, then $i_k$s will be independent of $\mu_i$ and the $p$-MAP estimator will be equivalent to the MAP estimator. Moreover, when  $K_i=2$ and $i_2=i^*$ for all $i$ with $\tau_{i,1}=\tau$, then the definitions of the $p$-MAP estimator and the $\tau$-MAP estimator coincide for CTMCs with a unique stationary maximum. 

\section{Main Results}\label{sec:main_results}
In this section, we present our main analytical results for MBF. We will first find the MBF of a martingale estimator and will later use its result to extend it to general estimators such as the $p$-MAP estimator.

\subsection{Martingale Estimator}
Let $\tilde{Z}(t)=(X(t),\hat{X}_M(t))$. Then, $\tilde{Z}(t)$ is a finite-state irreducible two-dimensional CTMC. Hence, its stationary distribution, denoted by  the column vector $\bm\psi=\{\psi_{i,j}\}_{i,j\in \mathcal{S}}$, exists. Let $Q_M$ be its generator matrix whose elements will be denoted by $Q_M[s_1,s_2]$, where $s_1,s_2\in S\cross S$. Then, $Q_M$ can be fully characterized as follows,
\begin{align}
    Q_M[(i,j),(k,l)]=\begin{cases}
        \mu_j,&\text{if}~k=l=i\neq j,\\
        q_{ik},&\text{if}~k\neq i ~\text{and}~l=j,\\
        -q_i+\mu_j, &\text{if}~k=i\neq j=l,\\
        -q_i,&\text{if}~k=l=i=j,\\
        0,& \text{otherwise}.
    \end{cases}
\end{align}
Then, $\bm \psi$ is the unique solution that satisfies the global balanced equations $\bm\psi^TQ_M=\bm{0}$ and $\bm\psi^T\bm{1}=1$ where $\bm{1}$ is a column vector of all ones of appropriate dimension. Let $\tilde{Q}_M$ be the transpose of matrix $Q_M$ with the last row replaced by a row of ones and let $v_{S^2}$ be a column vector of all zeros except at the last index. Then, $\bm\psi$ can be found as follows,
\begin{align}
    \bm\psi= \tilde{Q}_M^{-1}v_{S^2}.
\end{align}
Then, the MBF under a martingale estimator is given by,
\begin{align}
    \e[\Delta_M]=\sum_{i \in S}\psi_{i,i}.
\end{align}

\subsection{Non-martingale Estimators}
Now, we will find the MBF of the $p$-MAP and $\tau$-MAP estimators. We will first show that the MBF of a non-martingale estimator can be computed using $\bm\psi$ and the average time the estimator was fresh between sampling instances which we denote by $\e[F_{i,\mu_i}]$.

\begin{theorem}\label{thrm:fresh_mui}
    The MBF, $\e[\Delta]$, under a non-martingale estimator is given by,
    \begin{align}
        \e[\Delta]=\sum_{i=1}^S\mu_i\tilde{\pi}_i\e[F_{i,\mu_i}],
    \end{align}
    where $\tilde{\pi}_i=\sum_{j\in S}\psi_{j,i}$ is the proportion of time the martingale estimator was in state $i$.
\end{theorem}

Next, we explicitly find the expression for $\e[F_{i,\mu_i}]$ under any given estimator in Theorem \ref{thrm:F_i}.

\begin{theorem}\label{thrm:F_i}
    The expected portion of time  the estimator was fresh starting from state $i$,  will be given by,
    \begin{align}
        \e[F_{i,\mu_i}]=\int_{0}^\infty P_{i\hat{X}(t)}(t)e^{-\mu_it}\,dt.\label{eqn:Fint}
    \end{align}
\end{theorem}

\begin{corollary}\label{cor:pmap_mui}
    For a time-reversible CTMC, the MBF for the $p$-MAP estimator is given by,
    \begin{align}
    \e[\Delta_P]=\sum_{i=1}^S\sum_{j=1}^S\sum_{k=1}^{K_i}\tilde{a}_{i,j,k}(e^{-(d_j+\mu_i)\tau_{i,k-1}}-e^{-(d_j+\mu_i)\tau_{i,k}}),\label{eqn:pmap_mui}
    \end{align}
    where $\tilde{a}_{i,j,k}=\sqrt{\frac{\pi_{i_k}}{\pi_i}}\tilde{\pi}_iu_{ji}u_{ji_k}\frac{\mu_i}{\mu_i+d_j}$.
\end{corollary}

\begin{corollary}\label{cor:tmap_mui}
    The MBF under the $\tau$-MAP estimator is,
    \begin{align}
    \e[\Delta_\tau]=\sum_{i=1}^S\sum_{j=1}^S\frac{\mu_i}{d_j+\mu_i}\left(\tilde{\pi}_iu_{ji}^2-\tilde{b}_{i,j}e^{-(d_j+\mu_i)\tau}\right),\label{eqn:tmap_mui}
    \end{align}
    where $\tilde{b}_{i,j}=\tilde{\pi}_iu_{ji}^2-\sqrt{\frac{\pi_{i^*}}{\pi_i}}\tilde{\pi}_iu_{ji}u_{ji^*}$.
\end{corollary}

The proof of Corollary \ref{cor:pmap_mui} directly follows by evaluating the integral in \eqref{eqn:Fint}, and the proof of Corollary \ref{cor:tmap_mui} follows by setting $K_i=2$, $i_2=i^*$ and  $\tau_{i,1}=\tau$ in \eqref{eqn:pmap_mui}. We can further show that the the average sampling rate in this setting, which we denote by $\omega$, is given by $\omega=\sum_{i\in S}\tilde{\pi}_i\mu_i$. Note that the sampling rate, regardless of the estimator, is always equal to that of the martingale estimator as the sampling rates are adjusted only based on the most recent observation.

\begin{remark}
    If $\mu_i=\mu$, then \eqref{eqn:tmap_mui} reduces to the same expression in \cite{erlang_est} for the $\tau$-MAP estimator obtained using the Erlang approximation.
\end{remark}

\section{Optimal State-Dependent Sampling Policy}
In this setting, a natural question to ask is how one can optimally allocate the sampling rates while being subjected to a total sampling budget. This leads us to the following optimization problem,
\begin{maxi}
    {\mu_i}{\e[\Delta]}
    {\label{eqn:opt_state_dep}}
    {}
    \addConstraint{\sum_{i=1}^S \mu_i\tilde{\pi}_i}{\leq \Omega},
\end{maxi}
where $\Omega$ is the maximum average sampling rate allowed. Both the objective and the constraint of the above optimization problem are non-convex functions of $\mu_i$s. Therefore, finding closed-form solutions and analytically optimizing it is a difficult task. To circumnavigate these challenges, we formulate the problem as an SMDP and use a policy iteration algorithm to find the optimal sampling rates for each state.

\begin{remark}
    The results in this section are not limited to time-reversible CTMCs or $p$-MAP estimators. In fact, the SMDP framework used in this section is readily applicable  to any CTMC with any generic estimator as long as the integral in finding $\e[F_{i,\mu_i}]$ can be computed efficiently.
\end{remark}

Now, we construct the SMDP for solving the optimization problem in \eqref{eqn:opt_state_dep}. Instead of directly solving \eqref{eqn:opt_state_dep}, we will first convert it to an unconstrained problem with the use of a Langragian multiplier. Further, to reduce the size of the action space of the SMDP, we will restrict the feasible $\mu_i$s to be bounded from above and below. Moreover, for the SMDP, we enable the use of randomized policies by slightly deviating from the original problem which was restricted to the class of deterministic policies. This enables us to tackle a larger class of rate allocation policies generalizing our problem. Now, our optimization problem reduces to the following,
\begin{maxi}
    {\rho_l<\mu_i<\rho_u}{\e[\Delta]-\gamma\sum_{i=1}^S \mu_i\tilde{\pi}_i,}
    {\label{eqn:opt_uncons_lag}}
    {}
\end{maxi}
where $\rho_l<\Omega$ and is chosen very close to zero and $\rho_u$ is well above $\Omega$. The SMDP can be fully characterized using the tuple $(\mathcal{S,A,P,R,H)}$ defined below.
\begin{itemize}
    \item The \emph{state space} $\mathcal{S}=\{1,2,\dots,S\}$ is the state of our martingale estimator $\hat{X}_M(t)$.
    \item The \emph{action space} $\mathcal{A}=[\rho_l,\rho_u]$ is the set of feasible sampling rates.
    \item The \emph{transition function} $\mathcal{P}:\mathcal{S\cross A\cross S}\to [0,1]$ defines the transition probabilities between states based on the selected action. In particular, $\mathcal{P}(s,a,s')$ denotes the probability of transitioning to state $s'$ if the action $a$ was selected at state $s$. If $(s,a,s')=(i,\mu_i,j)$, then $\mathcal{P}(i,\mu_i,j)=\int_0^\infty P_{ij}(t)\mu_ie^{-\mu_it}\dd{t}$.
    \item The \emph{reward function} $\mathcal{R}:\mathcal{S\cross A}\to \mathds{R}$ defines the average reward obtained in a  state based on the action selected. In here, $\mathcal{R}(s,a)$ denotes the average reward obtained by selecting action $a$ in state $s$. When in state $i$, if we choose $\mu_i$ as the sampling rate, then $\mathcal{R}(i,\mu_i)=\e[F_{i,\mu_i}]-\gamma$.
    \item The \emph{sojourn times} $\mathcal{H}$ defines the average time the process stays in a particular state based on the action selected. In particular, $\mathcal{H}(s,a)$ denotes the average time the process stays in state $s$ if the action $a$ was selected. For example, if we choose rate $\mu_i$ when in state $i$, then $\mathcal{H}(i,\mu_i)=\frac{1}{\mu_i}$ since the martingale estimator changes only when a new sample is obtained.
\end{itemize}
Now, the optimal state-dependent policy for \eqref{eqn:opt_uncons_lag} will be a \emph{simple policy} (stationary and deterministic) \cite{Ross_CSMDP} and can be found using the policy iteration. Let us denote the optimal policy for the unconstrained problem by $\zeta^\gamma$ and its average sampling rate by $\omega^\gamma$.

Next, we need to translate our results from the unconstrained problem to the constrained problem. Since our action space limits the feasible sampling rates to be positive, the SMDP will be irreducible under any simple policy. Additionally, the simple policy $\mu_i=\Omega/2$ satisfies the sampling rate constraint. Therefore, from \cite{Ross_CSMDP}, we have that the optimal rate allocation policy for the constrained problem will be attained at either $\gamma=0$ or at some $\gamma>0$ which satisfies $\omega^\gamma=\Omega$. More specifically, if there exists a $\zeta^0$ such that $w^0\leq\Omega$, then $\zeta^0$ is the optimal policy for our constrained problem. Otherwise, if there exists some $\gamma>0$ such that $\omega^\gamma=\Omega$, then the optimal policy for the unconstrained problem with that particular $\gamma$ is optimal for the constrained problem. Finally, if both these conditions fail, then there exists some $\gamma$ such that $\lim_{\epsilon\downarrow0} \omega^{\gamma+\epsilon}=\Omega_0<\Omega<\Omega^0=\lim_{\epsilon\uparrow0} \omega^{\gamma+\epsilon}$. The existence of such a $\gamma$ is shown in \cite{Ross_MDP}. Moreover, \cite{Ross_MDP} shows that $w^\gamma$ is non-increasing in $\gamma$, and hence, a bisection search can be used to narrow down the $\gamma$. However, in this case, the optimal policy is not a simple policy but rather a \emph{semi-simple policy} (SSP) which means that there exists at most one state for which the optimal policy needs to be randomized between two actions and for every other state the action is deterministic.

To find the optimal SSP, let us define $\bar\zeta=\lim_{\epsilon\downarrow0}\zeta^{\gamma+\epsilon}$ and $\underline{\zeta}=\lim_{\epsilon\uparrow0}\zeta^{\gamma+\epsilon}$. Let $\bar\omega$ and $\underline{\omega}$ be the average sampling rates of the policies $\bar\zeta$ and $\underline{\zeta}$. Then, $\bar\omega=\Omega_0$ and $\underline{\omega}=\Omega^0$, and the optimal SSP can be obtained by randomizing between $\bar\zeta$ and $\underline{\zeta}$ as follows. Denote by $\zeta_s$, the action taken under policy $\zeta$ when in state $s$. Let $\tilde{\zeta}^k$ be a policy such that $\tilde{\zeta}^k_s=\bar{\zeta}_s$ for $s>k$ and $\tilde{\zeta}^k_s=\underline{\zeta}_s$ for $s\leq k$. Let $\tilde{\omega}^k$ be the average sampling rate of $\tilde{\zeta}^k$. Since $\tilde{\omega}^0=\Omega_0$ and $\tilde{\omega}^S=\Omega^0$, there exists a $k'$ such that  $\tilde{\omega}^{k'-1}\leq\Omega<\tilde{\omega}^{k'}$. Then, the optimal SSP $\zeta^*$ is given by $\zeta^*_s=\bar\zeta_s$ for $s>k'$, $\zeta^*_s=\underline{\zeta}_s$ for $s<k'$ and for $s=k'$, we randomize between $\bar\zeta_{k'}$ and $\underline{\zeta}_{k'}$ with some probability $p$, where $p$ is chosen such that average sampling rate of $\zeta^*$ is $\Omega$. 

\begin{remark}
    Our numerical results show that the optimal policy most likely turns out to be a simple policy whereas semi-simple policies may sometimes arise as an artifact of the discretization of the action space when finding the optimal actions in the policy iteration algorithm. In fact, when we further refined the action space, we noted that our algorithm resulted in simple policies.
\end{remark}

\section{Numerical Results}
In this section, we comprehensively evaluate the performance of the proposed structured estimators and compare our rate allocation policies with suitable benchmarks. We will use the terms rate allocation policies and  sampling policies interchangeably. In here, for the $p$-MAP estimator, we always assume that $\tau_{i,k}=\tau^*_{i,k}$ unless otherwise specified, and for the $\tau$-MAP estimator, $\tau=\tau^*$ is used. Moreover, to construct time-reversible CTMCs, we will also use  finite birth death chains (BDC) which are well-known to be time-reversible.

In the first experiment, we will use a finite BDC to compare our rate allocation policies. We denote by $\Delta_M^*$, $\Delta_\tau^*$ and $\Delta_p^*$, the MBF metric of the three estimators under their optimal sampling rates computed using the SMDP framework. We compare these optimal state-dependent sampling policies against a uniform rate allocation policy which allocates $\mu_i=\Omega$ for all $i$. This rate allocation policy is trivially contained within the feasible set of our original optimization problem for optimal state-dependent sampling policies, and hence it provides a natural lower bound for MBF curves under the optimal policy. Let $\Delta_M^U$, $\Delta_\tau^U$ and $\Delta_p^U$ denote the MBF of the three estimators under the uniform sampling policy.

\begin{figure}[t]
    \centering
    \includegraphics[width=0.85\columnwidth]{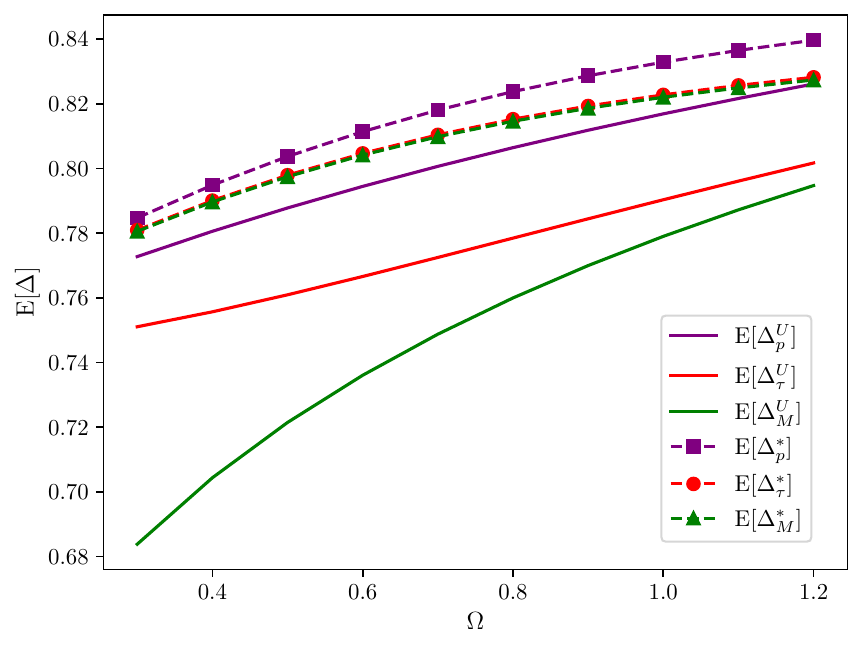}
    \caption{Variation of MBF with the sampling budget $\Omega$, under different state-dependent sampling policies, for a finite birth death chain (BDC) with $S=4$ states, with a unique stationary maximum.}
    \label{fig:var_omega}
\end{figure}

\begin{figure}[t]
    \captionsetup[subfigure]{aboveskip=1pt,belowskip=1pt}
    \centering
    \begin{subfigure}[b]{0.43\textwidth}
        \centering
        \includegraphics[width=\textwidth]{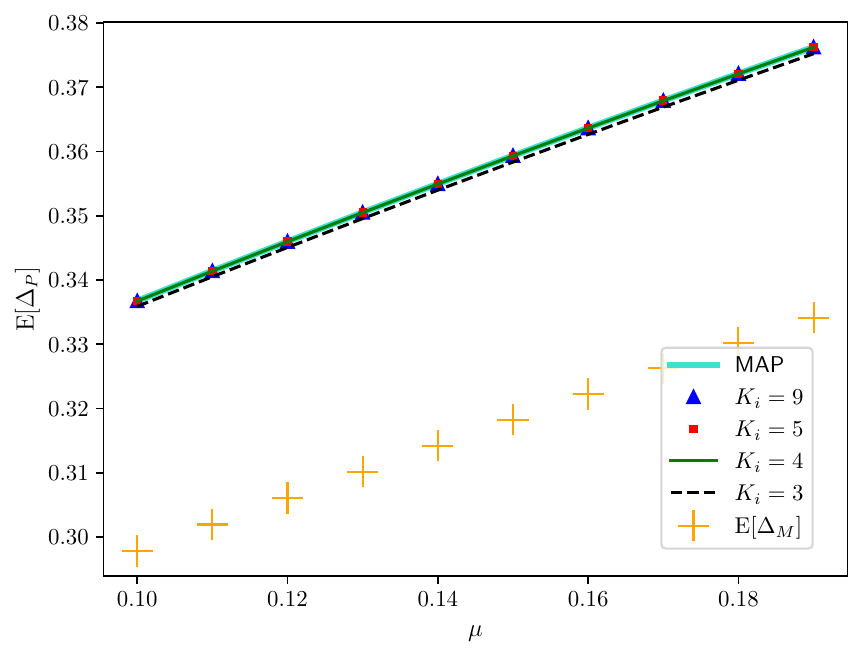}
        \vspace{-5mm}
        \caption{Using the first $K_i$ optimal transition points.}
        \label{fig:p_map_var}
    \end{subfigure}
    \medskip

    \begin{subfigure}[b]{0.43\textwidth}
        \centering
        \includegraphics[width=\textwidth]{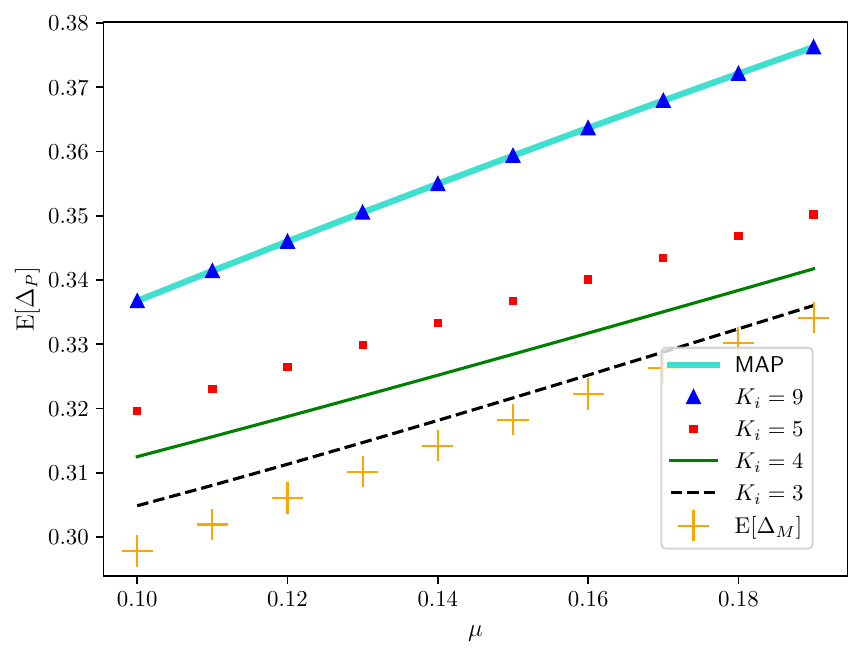}
        \caption{Periodically sampling the first nine optimal transition points.}
        \label{fig:p_map_var_periodic}
    \end{subfigure}
    \caption{Variation of MBF of the $p$-MAP estimator with the number of intermediate stages $K_i$ used to approximate an infinitely oscillating MAP estimator.}
    \label{fig:var_K_i}
\end{figure}

Fig.~\ref{fig:var_omega} illustrates the variation of MBF of the three estimators under the considered sampling policies for a BDC with a unique stationary maximum. We observe that, under any given sampling policy, the $p$-MAP estimator is always superior to the rest, and $\tau$-MAP estimator is only secondary to the $p$-MAP estimator. Moreover, we see that, under the optimal sampling policies, a significant improvement ($\sim15\%$ for martingale estimator and $\sim4\%$ for $\tau$-MAP estimator, at $\mu=0.3$) is observed compared to the uniform policy, highlighting the importance of a state-dependent sampling scheme. Also, when the optimal sampling policy is employed, the performance gap among the structured estimators narrows. This is mainly due to the fact that the optimal sampling policy will try to allocate higher rates to less probable states, and low rates to more probable states. Thus, when in a less probable state, we will quickly sample and acquire a fresher update whereas in more probable states, we are less likely to obtain new samples for longer durations. This improves the overall MBF by balancing out one of the main drawbacks of the martingale estimator, that is, having long sojourn times in less probable states.

Next, we compare how MBF varies with the number of stages $K_i$ used in the $p$-MAP estimator to approximate an infinitely oscillating MAP estimator. Here, we use a non-time-reversible ring structure similar to that of Fig.~\ref{fig:map_osc} to construct a CTMC with an infinitely oscillating MAP estimator. For this, we consider a CTMC with $S=4$ states placed in a ring structure where the transition rates alternate between two values. In particular, we transition from state $1$ to $2$ and from state $3$ to $4$ at a rate of $1$ whereas a rate of $0.75$ is used for transitions between states $2$ to $3$ and from state $4$ to $1$. No other transitions exist. Further, we assume that $\mu_i=\mu$, $\forall i$. Since the MAP estimator is infinitely oscillating, it will have infinitely many transition points $\tau_{i,k}^*$ given a starting state $i$. Ideally, if we use all these infinitely many transition points, our $p$-MAP estimator would resemble the MAP estimator exactly. In here, we will be using a subset of these optimal transition points for our $p$-MAP estimator. To construct these subsets, we have used two main methods. In the first method, for each starting state, we will use the first $K_i$ transition points of the MAP estimator to construct the $p$-MAP estimator, and in the next, we will periodically sample the first nine transition points of the MAP estimator at different periods leading to different $K_i$ values for our $p$-MAP estimator. Fig.~\ref{fig:p_map_var} and Fig.~\ref{fig:p_map_var_periodic} illustrate the MBF curves under these two methods, respectively. We observe that, in both cases, as the number of stages $K_i$ increases, the $p$-MAP estimator approaches the MAP estimator. Further, we see that having an exact approximation of the MAP estimator at the initial transition points is more important than the later transition points. In other words, from a freshness perspective, it is sufficient, if our estimators approximate the MAP estimator closely when the age of the estimator $\delta(t)$ is low, and more coarsely at higher ages.

\section{Conclusion}
In this work, we introduced a new estimator known as the $p$-MAP estimator which approximates the MAP estimator as a piecewise constant function. We provided an analytical framework to compute its freshness and showed that it provides a significant gain over the martingale estimator. Moreover, we showed that with an appropriate state-dependent sampling rate allocation scheme, freshness can be further enhanced.

\bibliographystyle{unsrt}
\bibliography{refs}

\end{document}